\documentclass[numbook, envcountsect, envcountsame, envcountreset, runningheads, smallextended]{svjour3}                     
\smartqed  
\usepackage{graphicx}
\usepackage{amsmath,amssymb}
\newcommand{\B}{{\cal B}}

\newcommand{\M}{{\cal M}}
\newcommand{\N}{{\cal N}}
\newcommand{\F}{{\cal F}}
\newcommand{\reals}{\mathbb{R}}
\newcommand{\ereals}{\overline{\mathbb{R}}}
\newcommand{\uball}{\mathbb{B}}
\newcommand{\diag}{\mathop{\rm diag}}
\newcommand{\eps}{\varepsilon}
\newcommand{\argmin}{\mathop{\rm argmin}\limits}
\newcommand{\dom}{\mathop{\rm dom}}
\newcommand{\rge}{\mathop{\rm rge}}
\newcommand{\upto}{{\raise 1pt \hbox{$\scriptstyle \,\nearrow\,$}}}
\newcommand{\downto}{{\raise 1pt \hbox{$\scriptstyle \,\searrow\,$}}}
\newcommand{\cl}{\mathop{\rm cl}\nolimits}

\newcommand{\FF}{(\F_t)_{t=0}^T}
\journalname{Finance and Stochastics}
\begin{document}

\title{Arbitrage and deflators in illiquid markets
}


\author{Teemu Pennanen}


\institute{T. Pennanen \at
              Department of Mathematics and Systems Analysis, Helsinki University of Technology, P.O.Box 1100, FI-02015 TKK, Finland \\
              \email{teemu.pennanen@tkk.fi}
}

\date{Received: date / Accepted: date}

\maketitle

\begin{abstract}
This paper presents a stochastic model for discrete-time trading in financial markets where trading costs are given by convex cost functions and portfolios are constrained by convex sets. The model does not assume the existence of a cash account/numeraire. In addition to classical frictionless markets and markets with transaction costs or bid-ask spreads, our framework covers markets with nonlinear illiquidity effects for large instantaneous trades. In the presence of nonlinearities, the classical notion of arbitrage turns out to have two equally meaningful generalizations, a marginal and a scalable one. We study their relations to state price deflators by analyzing two auxiliary market models describing the local and global behavior of the cost functions and constraints.
\end{abstract}

\section{Introduction}\label{intro}

When trading securities, marginal prices depend on the quantity traded. This is obvious already from the fact that different marginal prices are associated with purchases and sales. Marginal prices depend not only on the sign (buy/sell) but also on the size of the trade. When the trade affects the instantaneous marginal prices but not the marginal prices of subsequent trades, the dependence acts like a nonlinear transaction cost. Such short-term price impacts have been studied in several papers recently; see for example {\c{C}}etin, Jarrow and Protter~\cite{cjp4}, Rogers and Singh~\cite{rogsin6}, {\c{C}}etin and Rogers~\cite{cr7} and Astic and Touzi~\cite{at7} and their references. Short-term effects are different in nature from feedback effects where large trades have long-term price impacts that affect the marginal prices of transactions made at later times; see Kraus and Stoll~\cite{ks72} for comparison and empirical analysis of short- and long-term liquidity effects. Models for long-term price impacts have been developed e.g.\ in Platen and Schweizer~\cite{ps98}, Bank and Baum~\cite{bb4}. K\"uhn~\cite{kuh6}, Krokhmal and Uryasev~\cite{ku7}, Almgren and Chriss~\cite{ac} and Alfonsi, Schied and Schulz~\cite{ass} have proposed models that encompass both short and long run liquidity effects.

This paper presents a discrete time model for a general class of short-term liquidity costs. We model the total costs of purchases (positive or negative amounts) by random \em convex \em functions of the trade size. Convexity allows us to drop all assumptions about differentiability of the cost so that discontinuities in marginal prices can be modeled. This is essential e.g.\ in ordinary double auction markets, where marginal prices of market orders (instantaneous trades) are piecewise constant functions. It is necessary also if one wishes to cover models with transaction costs as e.g.\ in Jouini and Kallal~\cite{jk95a}.

The main observation of this paper is that in general convex models the notion of arbitrage turns out have two natural generalizations (see also \cite{pen7}, an earlier version of this paper). The first one is related to the possibility of producing something out of nothing and the second one to the possibility of producing arbitrarily much out of nothing. Accordingly, we introduce the conditions of \em no marginal arbitrage \em and \em no scalable arbitrage. \em In the case of sublinear models, as in classical market models or the models of \cite{jk95a} and \cite{km6}, the two notions coincide. In general, however, a market model can allow for marginal arbitrage while being free of scalable arbitrage. When there are no portfolio constraints, these notions of arbitrage are related to state price deflators that turn certain marginal price processes into martingales. Whereas marginal arbitrage is related to {\em market prices} associated with infinitesimal trades, scalable arbitrage is related to marginal prices contained in the closure of the whole range of possible marginal prices. In the presence of portfolio constraints, the martingale property is replaced by a more general one involving normal cones of the constraints much like in Pham and Touzi~\cite{pt99}, Napp~\cite{nap3}, Evstigneev, Sch{\"u}rger and Taksar~\cite{est4} and Rokhlin~\cite{rok5b,rok7} in the case of perfectly liquid markets with a cash account. 

Another, quite popular, approach to transaction costs is the currency market model of Kabanov~\cite{kab99}; see also Schachermayer~\cite{sch4}, Kabanov, R\'asonyi and Stricker~\cite{krs3} and their references. It treats proportional costs in a elegant way by specifying random solvency cones of portfolios that can be transformed into the zero portfolio at given time and state. This was generalized in Astic and Touzi~\cite{at7} to possibly nonconical solvency sets in the case of finite probability spaces. In these models, contingent claims and arbitrage are defined in terms of {\em physical delivery} (claims are portfolio-valued) as opposed to the more common {\em cash delivery}. Due to this difference and the fact that we allow for portfolio constraints, direct comparisons between existing results for the two classes of models are difficult even in the conical case. For example, the important issue of closedness of the set of claims that can be superhedged with zero cost is quite different if one looks at all claims rather than just those with cash delivery. Furthermore, the existence of portfolio constraints and the nonexistence of a cash-account/numeraire in our model brings up the important fact that, in practice, wealth cannot be transferred freely in time; see Dermody and Rockafellar~\cite{dr91,dr95} and Jouini and Napp~\cite{jn1}. This shifts attention to {\em contingent claim processes} that may give pay-outs not only at one date but possibly throughout the whole life time of the claim. Such claim processes are common in real markets. This suggests defining arbitrage in terms of contingent claim processes instead of static claims as in the classical perfectly liquid market model or those in \cite{kab99,sch4,at7,pen6}.

The rest of this paper is organized as follows. The market model is presented in Sections~\ref{sec:mm} and \ref{sec:pcp} together with some examples illustrating the differences between our model and existing ones. Section~\ref{sec:arb} defines the two notions of arbitrage and relates them to two conical market models. Section~\ref{sec:def} relates the notions of arbitrage to two kinds of deflators. Proofs of the main results are collected in the appendix.

\section{The market model}\label{sec:mm}

Most modern stock exchanges are based on the so called double auction mechanism to determine trades between market participants. In such an exchange, market participants submit offers to buy or sell shares within certain limits on the unit price and quantity. The trading system maintains a record, called the ``limit order book'', of all the offers that have {\em not} been offset by other offers. At any given time, the lowest unit price over all selling offers in the limit order book (the ``ask price'') is thus greater than the highest unit price over all buying offers (the ``bid price''). When buying in such a market, only a finite number of shares can be bought at the ask price and when buying more, one gets the second lowest price and so on. The \em marginal price \em for buying is thus a positive, nondecreasing, piecewise constant function of the number of shares bought. When selling shares, the situation is similar and the \em marginal price \em for selling is a positive, nonincreasing, piecewise constant function of the number of shares sold. 

Interpreting negative purchases as sales, we can incorporate the instantaneous margin\-al buying and selling prices into a single function $x\mapsto s(x)$ giving the marginal price for buying a positive or a negative number $x$ of shares at a fixed point in time. Since the bid price $\lim_{x\upto 0}s(x)$ is lower than the ask price $\lim_{x\downto 0}s(x)$, $s$ is a nonnegative nondecreasing function. If $x$ is greater than the total number of shares for sale we set $s(x)=+\infty$. The interpretation is that, at any given time, one cannot buy more than the total supply no matter how much one is willing to pay. On the other hand, if $x$ is less than the negative of the total demand we set $s(x)=0$ with the interpretation that one can not gain additional revenue by selling more than the total demand.

Given a marginal price function $s:\reals\to[0,+\infty]$ representing a limit order book, we can define the associated \em total cost function \em
\[
S(x):=\int_0^xs(w)dw,
\]
which gives the total cost of buying $x$ shares. The total cost $S:\reals\to\reals\cup\{+\infty\}$ associated with a nondecreasing marginal price $s:\reals\mapsto[0,+\infty]$ is an extended real-valued, lower semicontinuous convex function which vanishes at $0$; see Rockafellar~\cite[Theorem~24.2]{roc70a}. If $s$ happens to be finite everywhere, then by \cite[Theorem~10.1]{roc70a}, $S$ is not only lower semicontinuous but continuous.

In the above situation, the instantaneous marginal price is nonnegative and piecewise constant, or equivalently, the total cost is nondecreasing and polyhedral.
In the market model that we are about to present, the total cost is allowed to be a general lower semicontinuous convex function that vanishes at the origin. In particular, it allows negative marginal prices in situations where free disposal is not a valid assumption. Moreover, instead of a single asset we will allow for a finite set $J$ of assets and the total cost will be a function on the Euclidean space $\reals^J$ of portfolios.

Consider an intertemporal setting, where cost functions are observed over finite discrete time $t=0,\ldots,T$. Let $(\Omega,\F,P)$ be a probability space with a filtration $\FF$ describing the information available to an investor at each $t=0,\ldots,T$. For simplicity, we will assume that $\F_0$ is the trivial $\sigma$-algebra $\{\emptyset,\Omega\}$ and that each $\F_t$ is completed with respect to $P$. The Borel $\sigma$-algebra on $\reals^J$ will be denoted by $\B(\reals^J)$.

\begin{definition}\label{pp}
A \em convex cost process \em is a sequence $S=(S_t)_{t=0}^T$ of extended real-valued functions on $\reals^J\times\Omega$ such that for $t=0,\ldots,T$,
\begin{enumerate}
\item
the function $S_t(\cdot,\omega)$ is convex, lower semicontinuous and vanishes at $0$ for every $\omega\in\Omega$,
\item
$S_t$ is $\B(\reals^J)\otimes\F_t$-measurable.
\end{enumerate}
A cost process $S$ is said to be {\em nondecreasing, nonlinear, polyhedral, positively homogeneous, linear,} \ldots if the functions $S_t(\cdot,\omega)$ have the corresponding property for every $\omega\in\Omega$.
\end{definition}

The interpretation is that buying a portfolio $x_t\in\reals^J$ at time $t$ and state $\omega$ costs $S_t(x_t,\omega)$ units of cash. The measurability property implies that if the portfolio $x_t$ is $\F_t$-measurable then the cost $\omega\mapsto S_t(x_t(\omega),\omega)$ is also $\F_t$-measurable (see e.g.\ \cite[Proposition~14.28]{rw98}). This just means that the cost is known at the time of purchase. We pose no smoothness assumptions on the functions $S_t(\cdot,\omega)$.

The measurability property together with lower semicontinuity in Definition~\ref{pp}  mean that $S_t$ is an $\F_t$-measurable \em normal integrand \em in the sense of Rockafellar~\cite{roc68}; see Rockafellar and Wets~\cite[Chapter 14]{rw98} for an introduction to the theory of normal integrands. This has many important implications which will be used in the sequel.


Besides double auction markets as described earlier, Definition~\ref{pp} covers various more specific situations treated in the literature.

\begin{example}\label{ex:fl}
If $s_t$ is an $\reals^J$-valued $\F_t$-measurable price vector for each $t=0,\ldots,T$, then the functions
\[
S_t(x,\omega)=s_t(\omega)\cdot x
\]
define a linear cost process in the sense of Definition~\ref{pp}. This corresponds to a frictionless market (with possibly negative unit prices), where unlimited amounts of all assets can be bought or sold for prices $s_t$.
\end{example}

\begin{proof}
This is a special case of Example~\ref{ex:jk} below.
\qed\end{proof}

\begin{example}\label{ex:jk}
If $\overline s_t$ and $\underline s_t$ are $\reals^J$-valued $\F_t$-measurable price vectors with $\underline s_t\le\overline s_t$, then the functions
\[
S_t(x,\omega)=\sum_{j\in J} S_t^j(x^j,\omega),
\]
where
\[
S_t^j(x^j,\omega)=
\begin{cases}
\overline s_t^j(\omega)x^j & \text{if $x^j\ge 0$},\\
\underline s_t^j(\omega)x^j & \text{if $x^j\le 0$}
\end{cases}
\]
define a sublinear (i.e.\ convex and positively homogeneous) cost process in the sense of Definition~\ref{pp}. This corresponds to a market with transaction costs or bid-ask spreads, where unlimited amounts of all assets can be bought or sold for prices $\underline s_t$ and $\overline s_t$, respectively. This situation was studied in Jouini and Kallal~\cite{jk95a}. When $\underline s=\overline s$, one recovers Example~\ref{ex:fl}.
\end{example}

\begin{proof}
This is a special case of Example~\ref{ex:km} below.
\qed\end{proof}

\begin{example}\label{ex:km}
If $Z_t$ is an $\F_t$-measurable set-valued mapping from $\Omega$ to $\reals^J$, then the functions
\[
S_t(x,\omega)=\sup_{s\in Z_t(\omega)}s\cdot x
\]
define a sublinear cost process in the sense of Definition~\ref{pp}. This situation was studied in Kaval and Molchanov~\cite{km6} (in the case that the mappings $Z_t$ have convex compact values in the nonnegative orthant $\reals^J_+$). When $Z_t=[\underline s_t,\overline s_t]$
one recovers Example~\ref{ex:jk}.
\end{example}

\begin{proof}
The functions $S_t(\cdot,\omega)$ are clearly sublinear and vanish at $0$. By \cite[Example~14.51]{rw98}, $S_t(x,\omega)$ is also an $\F_t$-measurable normal integrand.
\qed\end{proof}

In Examples~\ref{ex:fl}, \ref{ex:jk} and \ref{ex:km} the cost process $S$ is positively homogeneous, which means that the size of a transaction has no effect on the unit price, only the direction matters. In that respect, the following model is more realistic.

\begin{example}\label{ex:cr}
If $s_t$ are $\reals^J_+$-valued $\F_t$-measurable vectors and $\varphi^j$ are lower semicontinuous convex functions on $\reals$ with $\varphi^j(0)=0$, then the functions
\[
S_t(x,\omega) = \sum_{j\in J}s_t^j(\omega)\varphi^j(x^j)
\]
define a convex cost process in the sense of Definition~\ref{pp}. The scalar case ($J$ is a singleton), with strictly positive $s$ and strictly convex, strictly increasing and differentiable $\varphi^j$ was studied in \c{C}etin and Rogers~\cite{cr7}.
\end{example}

\begin{proof}
See Proposition~14.44(d) and Corollary~14.46 of \cite{rw98}.
\qed\end{proof}


A potentially useful generalization of the above model is obtained by allowing the functions $\varphi^j$ to depend on $t$ and $\omega$. In fact, when it comes to modeling the dynamics of illiquidity the following turns out to be convenient; see \cite{mp8}.

\begin{example}\label{ex:mp}
If $s_t$ is an $\F_t$-measurable $\reals^J_+$-valued vector and $\varphi_t$ is an $\F_t$-measurable convex normal integrand on $\reals^J\times\Omega$ with $\varphi_t(0,\omega)=0$, then the functions
\[
S_t(x,\omega) = \varphi_t(M_t(\omega)x,\omega),
\]
where $M_t(\omega)=\diag(s_t(\omega))$ is the diagonal matrix with entries $s_t^j$, define a convex cost process in the sense of Definition~\ref{pp}. When $s=(s_t)_{t=0}^T$ is a ``market price'' process giving unit prices for infinitesimal trades, the numbers $s_t^jx^j$ give the ``market values'' of the traded amounts. In this case, the cost of illiquidity depends on the (pretrade) market value rather than on the quantity of the traded amount.
\end{example}

In addition to nonlinearities in prices, one often encounters portfolio constraints when trading in practice. As in Rokhlin~\cite{rok5b}, we will consider general convex portfolio constraints where at each $t=0,\ldots,T$ the portfolio $x_t$ is restricted to lie in a convex set $D_t$ which may depend on $\omega$.

\begin{definition}\label{pcp}
A {\em convex constraint process} is a sequence $D=(D_t)_{t=0}^T$ of set-valued mappings from $\Omega$ to $\reals^J$ such that for $t=0,\ldots,T$,
\begin{enumerate}
\item
$D_t(\omega)$ is closed, convex and $0\in D_t(\omega)$ for every $\omega\in\Omega$,
\item
the set-valued mapping $\omega\mapsto D_t(\omega)$ is $\F_t$-measurable.
\end{enumerate}
A constraint process $D$ is said to be {\em polyhedral, conical,} \ldots if the sets $D_t(\omega)$ have the corresponding property for every $\omega\in\Omega$.
\end{definition}

The classical case without constraints corresponds to $D_t(\omega)=\reals^J$ for every $\omega\in\Omega$ and $t=0,\ldots,T$.

\begin{example}\label{ex:ck}
Given a closed convex set $K\subset\reals^J$ containing the origin, the sets $D_t(\omega)=K$ define a (deterministic) convex constraint process in the sense of Definition~\ref{pcp}. This case has been studied e.g.\ by Cvitani{\'c} and Karatzas~\cite{ck92} and Pham and Touzi~\cite{pt99}.
\end{example}

In addition to obvious ``short selling'' constraints, the above model (even with conical $K$) can be used to model situations where one encounters different interest rates for lending and borrowing. Indeed, this can be done by introducing two separate ``cash accounts'' whose unit prices appreciate according to the two interest rates and restricting the investments in these assets to be nonnegative and nonpositive, respectively.

In the example above, the constraint process is deterministic. In the following example, a stochastic constraint process is constructed from stochastic matrices.

\begin{example}\label{ex:napp}
Given a closed convex set $K\subset\reals^L$ containing the origin and an $\FF$-adapted sequence $(M_t)_{t=0}^T$ of real $L\times J$ matrices, the sets
\[
D_t(\omega) = \{x\in\reals^J\,|\, M_t(\omega)x\in K\},
\]
define a convex constraint process in the sense of Definition~\ref{pcp}. The case with a polyhedral convex cone $K$ was studied by Napp~\cite{nap3} in connection with linear cost processes. 
\end{example}

\begin{proof}
It is easily checked that the sets $D_t(\omega)$ are closed and convex. The fact that each $D_t$ is $\F_t$-measurable follows, by \cite[Example~14.15]{rw98}, from $\F_t$-measurability of $M_t$.
\qed\end{proof}

If $M_t(\omega)=\diag(s_t(\omega))$ is the diagonal matrix with market prices of the traded assets on the diagonal, the above example corresponds to a situation where one has constraints on {\em market values} rather than on units held. A simple example that goes beyond the conical case studied in \cite{nap3} is when there are nonzero bounds on market values of investments.

\section{Portfolio and claim processes}\label{sec:pcp}

When wealth cannot be transfered freely in time (due to e.g.\ different interest rates for lending and borrowing) it is important to distinguish between payments that occur at different dates. A {\em (contingent) claim process} is an $\reals$-valued stochastic process $c=(c_t)_{t=0}^{T}$ that is adapted to $(\F_t)_{t=0}^T$. The value of $c_t$ is interpreted as the amount of cash the owner of the claim receives at time $t$. Such claim processes are common e.g.\ in insurance. The set of claim processes will be denoted by $\M$. 

A {\em portfolio process}, is an $\reals^J$-valued stochastic process $x=(x_t)_{t=0}^T$ that is adapted to $(\F_t)_{t=0}^T$. The vector $x_t$ is interpreted as a portfolio that is held over the period $[t,t+1)$. The set of portfolio processes will be denoted by $\N$. An $x\in\N$ superhedges a claim process $c\in\M$ with zero cost if it satisfies the {\em budget constraint}\footnote{Given an $\F_t$-measurable function $z_t:\Omega\to\reals^J$, $S_t(z_t)$ denotes the extended real-valued random variable $\omega\mapsto S_t(z_t(\omega),\omega)$. By \cite[Proposition~14.28]{rw98}, $S_t(z_t)$ is $\F_t$-measurable whenever $z_t$ is $\F_t$-measurable.} 
\[
S_t(x_t-x_{t-1}) + c_t \le 0\quad P\text{-a.s.}\quad t=0,\ldots,T,
\]
and $x_T=0$. Here and in what follows, we always set $x_{-1}=0$. The above is a numeraire-free way of writing the superhedging property; see Example~\ref{ex:numeraire} below. In the case of a stock exchange, the interpretation is that the portfolio is updated by market orders in a way that allows for delivering the claim without any investments over time. In particular, when $c_t$ is strictly positive, the cost $S_t(x_t-x_{t-1})$ of updating the portfolio from $x_{t-1}$ to $x_t$ has to be strictly negative (market order of $x_{t-1}-x_t$ involves more selling than buying). At the terminal date, we require that everything is liquidated so the budget constraint becomes $S_T(-x_{T-1})+c_T\le 0$. 

The set of all claim processes that can be superhedged with zero cost under constraints $D$ will be denoted by $C(S,D)$. That is, 
\[
C(S,D) = \{c\in\M\,|\,\exists x\in\N_0:\ x_t\in D_t,\ S_t(\Delta x_t)+c_t\le 0,\ t=0,\ldots,T\},
\]
where $\N_0=\{x\in\N\,|\,x_T=0\}$. In other words, $C(S,D)$ consists of the claim processes that are freely available in the market. 

If it is assumed that a numeraire does exist, the above can be written in a more traditional form.


\begin{example}[Numeraire and stochastic integrals]\label{ex:numeraire}
Assume that there is a perfectly liquid asset, say $0\in J$, such that the cost functions can be written as 
\[
S_t(x,\omega) = s_t^0(\omega)x^0 + \tilde S_t(\tilde x,\omega),
\]
where $s^0$ is a strictly positive scalar process, $x=(x^0,\tilde x)$ and $\tilde S$ is a cost process for the remaining assets $\tilde J=J\setminus\{0\}$. Dividing the budget constraint by $s^0_t$, we can write it as
\[
x_t^0-x^0_{t-1} + \hat S_t(\tilde x_t-\tilde x_{t-1}) + \hat c_t \le 0\quad t=0,\ldots,T,
\]
where $(x^0_{-1},\tilde x_{-1})=(x^0_T,\tilde x_T)=0$ and 
\[
\hat S_t=\frac{1}{s_ t^0}\tilde S_t\quad\text{and}\quad \hat c_t=\frac{1}{s_ t^0}c_t
\]
are the cost function and the claim, respectively, in units of the numeraire. 

Given the $\tilde J$-part, $\tilde x=(\tilde x_t)_{t=0}^T$, of a portfolio process, we can define the numeraire part recursively by
\[
x_t^0 = x^0_{t-1} - \hat S_t(\tilde x_t-\tilde x_{t-1}) - \hat c_t \quad t=0,\ldots,T-1,
\]
so that the budget constraint holds as an equality for $t=1,\ldots,T-1$ and
\[
x^0_{T-1} = - \sum_{t=0}^{T-1}\hat S_t(\tilde x_t-\tilde x_{t-1}) - \sum_{t=0}^{T-1}\hat c_t.
\]
For $T$, the budget constraint thus becomes
\[
\sum_{t=0}^T\hat S_t(\tilde x_t-\tilde x_{t-1}) + \sum_{t=0}^T\hat c_t \le 0
\]
and we have
\[
C(S,D) = \{c\in\M\,|\,\exists\tilde x:\ \sum_{t=0}^T c_t/s^0_t \le -\sum_{t=0}^T\hat S_t(\tilde x_t-\tilde x_{t-1})\}.
\]
Thus, when a numeraire exists, hedging of a claim process can be reduced to hedging cumulated claims at the terminal date. If moreover, the cost process $\hat S$ is linear, i.e.\ $\hat S_t(\tilde x)=\hat s_t\cdot\tilde x$ we have
\[
\sum_{t=0}^T\hat S_t(\tilde x_t-\tilde x_{t-1}) = \sum_{t=0}^T\hat s_t\cdot(\tilde x_t-\tilde x_{t-1}) = -\sum_{t=0}^{T-1}\tilde x_t\cdot(\hat s_{t+1}-\hat s_t)
\]
and 
\[
C(S,D) = \{c\in\M\,|\,\exists\tilde x:\ \sum_{t=0}^T c_t/s^0_t \le \sum_{t=0}^{T-1}\tilde x_t\cdot(\hat s_{t+1}-\hat s_t)\}.
\]
Thus, in the classical linear model with a numeraire, the hedging condition can be written in terms of a stochastic integral as is often done in mathematical finance.
\end{example}

\begin{remark}[Market values]\label{rem:mv}
Instead of describing portfolios in terms of units, one could describe them, as in Kabanov~\cite{kab99}, in terms of ``market values''. Assume that 
\[
S_t(x,\omega) = \varphi_t(M_t(\omega)x,\omega)
\]
with $M_t(\omega)=\diag(s_t(\omega))$ as in Example~\ref{ex:mp}. If $s_t^j$ is the market price of asset $j\in J$, then the market value of $x^j$ units of the asset is $s_t^jx_t^j$. Making the change of variables $h_t^j(\omega) := s_t^j(\omega)x_t^j(\omega)$ and assuming that $s^j$ are strictly positive, we can write the budget constraint as
\[
\varphi_t(h_t-R_th_{t-1}) + c_t \le 0,
\]
where $R_t$ is the diagonal matrix with ``market returns'' $s_t^j/s_{t-1}^j$ on the diagonal.
\end{remark}

\begin{remark}[Physical delivery]\label{rem:cur}
In this paper, we study claim processes with {\em cash-delivery} but one could also study claim processes with {\em physical delivery} whose pay-outs are random portfolios. One could say that a portfolio process $x$ superhedges an $\reals^J$-valued claim process $c$ with zero initial cost if 
\[
S_t(\Delta x_t + c_t)\le 0 \quad P\text{-a.s.}\quad t=0,\ldots,T,
\]
where $x_{-1}=x_T=0$. Defining the $\F_t$-measurable closed convex set
\[
K_t(\omega):=\{x\in\reals^J\,|\, S_t(x,\omega)\le 0\}
\]
of portfolios available for free, the above budget constraint can be written
\[
\Delta x_t + c_t \in K_t\quad P\text{-a.s.}\quad t=0,\ldots,T.
\]
If there are no portfolio constraints, then much as in Example~\ref{ex:numeraire}, this could be written in terms of a static $\reals^J$-valued claim with maturity $T$. This would be similar to \cite{kab99,sch4,krs3,at7}. In the presence of portfolio constraints, claim processes cannot be reduced to claims with single payout date.

In perfectly liquid markets without portfolio constraints, a claim with physical delivery reduces to a claim with cash-delivery. Conversely, in the presence of a cash account, a claim $c$ with cash delivery can be treated as a claim with physical delivery. In general illiquid markets without a cash account and with portfolio constraints, contingent claims with physical delivery and those with cash-delivery are genuinely different objects.
\end{remark}

\section{Two kinds of arbitrage}\label{sec:arb}

Consider a market described by a convex cost process $S$ and a convex constraint process $D$. We will say that $S$ and $D$ satisfy the {\em no arbitrage} condition if there are no nonzero nonnegative claim processes which can be superhedged with zero cost by a feasible portfolio process. The no arbitrage condition can be written as
\[
C(S,D)\cap\M_+=\{0\},
\]
where $\M_+$ denotes the set of nonnegative claim processes. This is analogous to B\"uhlmann, Delbaen, Embrechts and Shiryaev~\cite[Definition 2.3]{bdes98} and Jouini and Napp~\cite[Definition~2.3]{jn1} but there is an essential difference in that $C(S,D)$ need not be a cone.

\begin{lemma}\label{lem:convex}
The set $C(S,D)$ is convex and contains all nonpositive claim processes. If $S$ is sublinear and $D$ is conical, then $C(S,D)$ is a cone.
\end{lemma}

\begin{proof}
Since $S_t(0)=0$, $C(S,D)$ contains the nonpositive claims. The rest will follow from the facts that $C(S,D)$ is the image of the set 
\[
E = \{(x,c)\in\N_0\times\M\,|\, x_t\in D_t,\ S_t(\Delta x_t)+c_t\le 0,\ t=0,\ldots,T\},
\]
under the projection $(x,c)\mapsto c$ and that the set $E$ is convex (cone) whenever $S$ is convex (sublinear) and $D$ is convex (and conical). To verify the convexity of $E$ let $(x^i,c^i)\in E$ and $\alpha^i>0$ such that $\alpha^1+\alpha^2=1$. By convexity of $D$, $\alpha^1x^1_t+\alpha^2x^2_t\in D_t$ and by convexity of $S$
\begin{align*}
S_t[\Delta (\alpha^1x^1+\alpha^2x^2)_t] + \alpha^1c^1_t + \alpha^2c^2_t
&= S_t[\alpha^1\Delta x^1_t + \alpha^2\Delta x^2_t] + \alpha^1c^1_t + \alpha^2c^2_t\\
&\le \alpha^1S_t(\Delta x^1_t) + \alpha^2S_t(\Delta x^2_t)  + \alpha^1c^1_t + \alpha^2c^2_t\\
&\le\alpha^1[S_t(\Delta x^1_t) +c^1_t] + \alpha^2[S_t(\Delta x^2_t) + c^2_t]\\
&\le 0.
\end{align*}
Thus $(\alpha^1x^1+\alpha^2x^2,\alpha^1c^1 + \alpha^2c^2)\in E$, so $E$ is convex. If $D$ is conical and $S$ is sublinear, the same argument works with arbitrary $\alpha^i>0$ which implies that $E$ is a cone.
\qed\end{proof}

In the classical linear model, or more generally, when $S$ is sublinear and $D$ is conical, the set $C(S,D)\cap\M_+$ is a cone, which means that arbitrage opportunities (if any) can be scaled by arbitrary positive numbers to yield arbitrarily ``large'' arbitrage opportunities. In general illiquid markets, this is no longer true and one can distinguish between two kinds of arbitrage opportunities: the original ones defined as above and those that can be scaled by arbitrary positive numbers without leaving the set $C(S,D)$.

\begin{definition}
A cost process $S$ and a constraint process $D$ satisfy the condition of {\em no scalable arbitrage} if
\[
\left(\bigcap_{\alpha>0}\alpha C(S,D)\right)\cap \M_+ = \{0\}.
\]
\end{definition}

Obviously, the no arbitrage condition implies the no scalable arbitrage condition and when $C(S,D)$ is a cone, the two coincide. In general, however, a market model may allow for arbitrage but still be free of scalable arbitrage. A simple condition guaranteeing the no scalable arbitrage condition is that
\[
\inf_{x\in\reals^J}S_t(x)>-\infty\quad P\text{-}a.s.,\quad t=0,\ldots,T.
\]
Indeed, the elements of $C(S,D)$ are uniformly bounded from above by the function $(\omega,t)\mapsto -\inf_xS_t(x,\omega)$ so if this is finite, $\bigcap_{\alpha>0}\alpha C$ is contained in $\M_-$. The condition $\inf S_t(x)>-\infty$ means that the revenue one can generate by an instantaneous transaction at given time and state is bounded from above. In the case of double auction markets, it simply corresponds to the fact that the ``bid-side'' of the limit order book has finite depth. Another condition excluding scalable arbitrage opportunities is that the sets $D_t$ be almost surely bounded for every $t=0,\ldots,T$.

Since $\M_+$ is a cone, the no arbitrage condition $C(S,D)\cap\M_+=\{0\}$ is equivalent to the seemingly stronger condition
\[
\left(\bigcup_{\alpha>0}\alpha C(S,D)\right)\cap \M_+ = \{0\},
\]
as is easily verified. 
The two sets
\[
\bigcup_{\alpha>0}\alpha C(S,D)\quad\text{and}\quad \bigcap_{\alpha>0}\alpha C(S,D)
\]
are convex cones and they both coincide with $C(S,D)$ when $C(S,D)$ is a cone. The two cones can be described in terms of two auxiliary market models with a sublinear costs and conical constraints. This will be used in the derivation of our main results below. An alternative approach can be found in \cite{pen7}. 

Given an $\alpha>0$, it is easily checked that 
\[
(\alpha\star S)_t(x,\omega):=\alpha S_t(\alpha^{-1} x,\omega).
\]
defines a convex cost process in the sense of Definition~\ref{pp} and that 
\[
(\alpha D)_t(\omega):=\alpha D_t(\omega)
\]
defines a convex constraint process in the sense of Definition~\ref{pcp}. With this notation, we have 
%
\begin{align*}
\alpha C(S,D) &= \{\alpha c\,|\,\exists x:\ x_t\in D_t,\ S_t(\Delta x_t)+c_t\le 0\}\\
&= \{c'\,|\,\exists x:\ x_t\in D_t,\ \alpha S_t(\Delta x_t)+c'_t\le 0\}\\
&= \{c'\,|\,\exists x':\ x'_t\in\alpha D_t,\ \alpha S_t\left(\frac{\Delta x'_t}{\alpha}\right)+c'_t\le 0\}\\
&= C(\alpha\star S,\alpha D).
\end{align*}
If $S$ is positively homogeneous, we simply have $\alpha\star S=S$, but in the general convex case, $\alpha\star S$ decreases as $\alpha$ increases; see \cite[Theorem~23.1]{roc70a}. In particular, pointwise limits of $\alpha\star S$ exist when $\alpha$ tends to zero or infinity. The lower limit, $\inf_{\alpha >0}\alpha\star S_t(x,\omega)$ is nothing but the directional derivative of $S_t(\cdot,\omega)$ at the origin. Its lower semicontinuous hull will be denoted by
\[
S_t'(x,\omega) := \liminf_{x'\to x}\inf_{\alpha >0}\alpha\star S_t(x',\omega).
\]
By \cite[Theorem~23.4]{roc70a}, the directional derivative is automatically lower semicontinuous when the origin is in the relative interior of $\dom S_t(\cdot,\omega):=\{x\in\reals^J\,|\, S_t(x,\omega)<\infty\}$. The upper limit
\[
S_t^\infty(x,\omega) := \sup_{\alpha>0}\alpha\star S_t(x,\omega)
\]
is automatically lower semicontinuous, by lower semicontinuity of $\alpha\star S_t(\cdot,\omega)$ (which in turn follows from that of $S_t(\cdot,\omega)$). Whereas $S_t'$ describes the local behavior of $S_t$ near the origin, $S_t^\infty$ describes the behavior of $S_t$ infinitely far from it. In the terminology of variational analysis, $S_t'(\cdot,\omega)$ is the {\em subderivative} of $S_t(\cdot,\omega)$ at the origin, whereas $S_t^\infty(\cdot,\omega)$ is the {\em horizon function} of $S_t(\cdot,\omega)$; see Theorem~3.21 and Proposition~8.21 of \cite{rw98}. If $S$ is sublinear, then $S_t'=S_t^\infty=S_t$. In general, we have the following.

\begin{proposition}
Let $S$ be a convex cost process. The sequences $S'=(S_t')_{t=0}^T$ and $S^\infty=(S_t^\infty)_{t=0}^T$ define sublinear cost processes in the sense of Definition~\ref{pp}. The process $S'$ is the greatest sublinear cost process less than $S$ and $S^\infty$ is the least sublinear cost process greater than $S$. 
\end{proposition}

\begin{proof}
The properties in the first condition of Definition~\ref{pp} follow from convexity; see Proposition~8.21 and Theorem~3.21 of \cite{rw98}. The measurability properties follow from Theorem~14.56 and Exercise~14.54 of \cite{rw98}.
\qed\end{proof}

Analogously, if $D$ is conical, we have $\alpha D=D$, but in the general convex case, $\alpha D$ gets larger when $\alpha$ increases. We define
\begin{align*}
D_t'(\omega) &= \cl\bigcup_{\alpha>0}\alpha D_t(\omega),\\
D_t^\infty(\omega) &= \bigcap_{\alpha>0}\alpha D_t(\omega),
\end{align*}
where the closure is taken $\omega$-wise in $\reals^J$. Whereas $D_t'(\omega)$ describes the local behavior of $D_t(\omega)$ near the origin, $D_t^\infty(\omega)$ describes the behavior of $D_t(\omega)$ infinitely far from it. In the terminology of variational analysis, $D_t'(\omega)$ is the \em tangent \em cone of $D_t(\omega)$ at $0$ and $D_t^\infty(\omega)$ is the \em horizon cone \em of $D_t(\omega)$; see Theorems~3.6 and 6.9 of \cite{rw98}. When $D_t(\omega)$ is polyhedral, then its positive hull $\bigcup_{\alpha>0}\alpha D_t(\omega)$ is automatically closed and the closure operation in $D'$ is superfluous. In general, however, the positive hull is not closed. If $D$ is conical, then $D_t'=D_t^\infty=D_t$. In general, we have the following.

\begin{proposition}
Let $D$ be a convex constraint process. The sequences $D'=(D_t')_{t=0}^T$ and $D^\infty=(D_t^\infty)_{t=0}^T$ define convex conical constraint processes in the sense of Definition~\ref{pcp}. The process $D'$ is the smallest conical constraint process containing $D$ and $D^\infty$ is the largest conical constraint process contained in $D$.
\end{proposition}

\begin{proof}
The properties in the first condition of Definition~\ref{pcp} are easy consequences of convexity; see Theorems~3.6 and 6.9 of \cite{rw98}. The measurability properties come from Exercise~14.21 and Theorem~14.26 of \cite{rw98}.
\qed\end{proof}

When the cost process $S$ is finite-valued (i.e.\ $S_t(x,\omega)<\infty$ for every $t$, $\omega$ and $x\in\reals^J$), we get the following estimates for the two cones involved in the no arbitrage conditions. Here and in what follows, $\uball$ denotes the Euclidean unit ball of $\reals^J$.

\begin{proposition}\label{prop:cones}
Assume that $S$ is finite-valued. Then
\[
\bigcup_{\alpha>0}\alpha C(S,D) \subset C(S',D') \subset \cl\bigcup_{\alpha>0}\alpha C(S,D),
\]
where the closure is taken in terms of convergence in probability. If there is an $a\in L^0_+$ such that $S\ge S^\infty-a$ and $D\subset D^\infty+a\uball$, then
\[
C(S^\infty,D^\infty) \subset \bigcap_{\alpha>0}\alpha C(S,D) \subset \cl C(S^\infty,D^\infty).
\]
\end{proposition}

\begin{proof}
See the appendix.
\qed\end{proof}

By the first part of Proposition~\ref{prop:cones}, the closure of $C(S',D')$ equals the tangent cone of $C(S,D)$ at the origin. The term ``arbitrage'' is usually replaced by the term ``free lunch'' when the condition is strengthened with the closure operation. Marginal and scalable notions of free lunch were the focus in the earlier papers \cite{pen6,pen7} on nonlinear convex cost functions; see also \cite{jn1} where a general class of conical models were studied as well as \cite{rok5b} on convex constraints and linear cost functions with a cash account. As illustrated in \cite[Example~1]{rok5b}, the closure operation is not superfluous in nonpolyhedral models even in the case of finite $\Omega$. On the other hand, it may happen that $C(S,D)$ is closed but its positive hull is not, even when $\Omega$ is finite; see \cite[p.~439]{rok5} for an example with linear costs and convex constraints. The following illustrates the phenomenon with nonlinear costs and without constraints.

\begin{example}\label{ex:tangent}
Let $T=1$, $J=\{1\}$ (just one asset), $S_0(x,\omega)=x$, $S_1(x,\omega)=e^x-1$ and $D_0(\omega)=D_1(\omega)=\reals$ for every $\omega\in\Omega$. Then
\begin{align*}
C(S,D) &= \{(c_0,c_1)\,|\,\exists x_0\in\reals:\ S_0(x_0)+c_0\le 0,\ S_1(-x_0)+c_1\le 0\}\\
&= \{(c_0,c_1)\,|\,\exists x_0\in\reals:\ x_0+c_0\le 0,\ e^{-x_0}-1+c_1\le 0\}\\
&= \{(c_0,c_1)\,|\, e^{c_0}-1+c_1\le 0\},
\end{align*}
so that
\[
\alpha C(S,D) = \{(c_1,c_1)\,|\, c_1\le\alpha(1-e^{c_0/\alpha})\}.
\]
As $\alpha$ increases, this set converges towards the set $\{(c_1,c_1)\,|\, c_1\le -c_0\}$ but it only intersects the line $c_1=-c_0$ at the origin. We thus get
\[
\bigcup_{\alpha>0}\alpha C(S,D)= \{(c_0,c_1)\,|\, c_1< -c_0\}\cup\{(0,0)\}
\]
which is not closed. Note that this model satisfies the no arbitrage condition.
\end{example}

When $S$ is sublinear and $D$ is conical, we have $S=S^\infty$ and $D=D^\infty$ so the extra conditions in the second part of Proposition~\ref{prop:cones} are automatically satisfied. More generally, by Corollary 9.1.2 and Theorem~8.4 of \cite{roc70a}, the condition $D\subset D^\infty+a\uball$ is satisfied in particular when $D_t=K_t+B_t$ for an $\F_t$-measurable closed convex cone $K_t$ and an $\F_t$-measurable almost surely bounded closed convex set $B_t$.

\section{Two kinds of deflators}\label{sec:def}

Given a convex cost process $S=(S_t)_{t=0}^T$ and an $x\in\reals^J$, the set of {\em subgradients}
\[
\partial S_t(x,\omega) := \{v\in\reals^J\,|\, S_t(x',\omega)\ge S_t(x,\omega) + v\cdot (x'-x)\quad\forall x'\in\reals^J\}
\]
is a closed convex set which is $\F_t$-measurable in $\omega$; see \cite[Theorem~14.56]{rw98}. The random $\F_t$-measurable set $\partial S_t(x)$ gives the set of \em marginal prices \em at~$x$. In particular, $\partial S_t(0)$ can be viewed as the set of {\em market prices} which give the marginal prices associated with infinitesimal trades in a market described by $S$. In the scalar case (when $J$ is a singleton), $\partial S_t(0)$ is the closed interval between the bid and ask prices, i.e.\ left and right directional derivatives of $S_t$ at the origin. If $S_t(x,\omega)$ happens to be differentiable at a point $x$, we have $\partial S_t(x,\omega)=\{\nabla S_t(x,\omega)\}$.

Given a convex constraint process $D=(D_t)_{t=0}^T$ and an $x\in\reals^J$, the {\em normal cone}
\[
N_{D_t(\omega)}(x):= 
\begin{cases}
\{v\in\reals^J\,|\, v\cdot (x'-x)\le 0\quad\forall x'\in D_t(\omega)\} & \text{if $x\in D_t(\omega)$},\\
\emptyset & \text{otherwise}
\end{cases}
\]
is a closed convex set which is $\F_t$-measurable in $\omega$; see \cite[Theorem~14.26]{rw98}. The random $\F_t$-measurable set $N_{D_t}(x)$ gives the set of price vectors $v\in\reals^J$ such that the portfolio $x\in\reals^J$ maximizes the value $v\cdot x'$ over all $x'\in D_t$. In particular, $N_{D_t}(0)$ gives the set of price vectors $v\in\reals^J$ such that the zero portfolio maximizes $v\cdot x$ over $D_t$. When $D_t(\omega)=\reals^J$ (no portfolio constraints), we have $N_{D_t(\omega)}(x)=\{0\}$ for every $x\in\reals^J$.

\begin{definition}\label{def:def1}
A nonnegative stochastic process $y$ is a {\em market price deflator} for $S$ and $D$ if there is a market price process $s\in\partial S(0)$ such that $ys$ is integrable and
\[
E[y_{t+1}s_{t+1}\,|\,\F_t]-y_ts_t\in N_{D_t}(0)
\]
$P$-almost surely for $t=0,\ldots,T$.
\end{definition}

When $D_t\equiv\reals^J$ (no portfolio constraints), we have $N_{D_t}=\{0\}$, so market price deflators are the nonnegative processes that turn some market price process into a martingale. In the presence of a numeraire, nonzero market price deflators correspond to absolutely continuous probability measures under which discounted market price process is a martingale. When $D_t(\omega)=\reals^J_+$, we have $N_{D_t(\omega)}(0)=\reals^J_-$ and the second inclusion means that $ys$ is a super-martingale. The general normal cone condition in Definition~\ref{def:def1} is essentially the same as the one obtained in Rokhlin~\cite{rok2} in the case of finite probability spaces and in the presence of a cash account; see also \cite{rok5b} for general probability spaces and linear cost process with a cash account

When the cost process $S$ happens to be smooth at the origin, market price deflators are the nonnegative processes $y$ such that
\[
y_t\nabla S_t(0)-E[y_{t+1}\nabla S_{t+1}(0)\,|\,\F_t]\in N_{D_t}(0)
\]
This resembles the martingale condition in Theorem~3.2 of {\c{C}}etin, Jarrow and Protter~\cite{cjp4} which says that (in a market with a cash account and without portfolio constraints) the value of the {\em supply curve} at the origin is a martingale under a measure equivalent to $P$. However, the supply curve of \cite{cjp4} is not the same as the marginal price. Indeed, the supply curve of \cite{cjp4} gives ``the stock price, per share, at time $t\in [0, T]$ that the trader pays/receives for an order of size $x\in\reals$''; see \cite[Section~2.1]{cjp4}. In our notation, the supply curve of \cite{cjp4} thus corresponds to the function $x\mapsto S_t(x)/x$ which agrees with the marginal price $\nabla S_t(x)$ in the limit $x\to 0$ (if $S_t(x)$ is smooth at the origin) but is different in general.

\begin{theorem}\label{thm:def1}
If there is a strictly positive market price deflator, then
\[
C(S',D')\cap\M_+=\{0\}.
\]
On the other hand, if $S'$ is finite-valued and 
\[
\cl C(S',D')\cap\M_+=\{0\},
\]
then there is a strictly positive market price deflator. Moreover, in this case, the market price deflator can be chosen bounded.
\end{theorem}

\begin{proof}
See the appendix.
\qed\end{proof}

Combining Theorem~\ref{thm:def1} with Proposition~\ref{prop:cones} we see that when $S$ is finite-valued, a bounded strictly positive market price deflator exists if
\[
\left[\cl\bigcup_{\alpha>0}\alpha C(S,D)\right]\cap\M_+=\{0\},
\]
which might be called the condition of {\em no marginal arbitrage}. Recall that when $S$ is sublinear and $D$ is conical, we have $C(S,D)=\bigcup_{\alpha>0}\alpha C(S,D)$. Furthermore, it is well-known that in the classical linear model with a cash account and without constraints, the no arbitrage condition implies that $C(S,D)$ is closed; see Schachermayer~\cite{sch92}. In nonpolyhedral models (even with finite $\Omega$) however, the closure operation is not superfluous; see \cite[Example~1]{rok5b}.

Whereas $\partial S_t(0)$ gives the set of market prices, the random $\F_t$-measurable set $\rge\partial S_t:=\bigcup_{x\in\reals^J}\partial S_t(x)$ gives the set of all possible marginal prices one may face when trading at time $t$ in a market described by $S$. Similarly, the random $\F_t$-measurable set $\rge N_{D_t}:=\bigcup_{x\in\reals^J}N_{D_t}(x)$ gives all the possible normal vectors associated with a constraint process $D$ at time $t$.

\begin{definition}\label{def:def2}
A nonnegative stochastic process $y$ is a {\em marginal price deflator} for $S$ and $D$ if there is a price process $s\in\cl\rge\partial S$ such that $ys$ is integrable and
\[
E[y_{t+1}s_{t+1}\,|\,\F_t]-y_ts_t\in\cl\rge N_{D_t}
\]
$P$-almost surely for $t=0,\ldots,T$.
\end{definition}

When $S$ is sublinear and $D$ is conical, we have $S=S'=S^\infty$, $D=D'=D^\infty$, $\rge\partial S_t=\partial S_t(0)$ and $\rge N_{D_t} = N_{D_t}(0)$ (which are closed sets), so that marginal price deflators coincide with market price deflators. If $D_t(\omega)\equiv\reals^J$ (no portfolio constraints), we have $\rge N_{D_t(\omega)}=\{0\}$ so the marginal price deflators are the nonnegative processes $y=(y_t)_{t=0}^T$ such that there is some marginal price process $s\in\cl\rge\partial S_t$ such that the deflated price process $ys=(y_ts_t)_{t=0}^T$ is a martingale. 

When $S$ is polyhedral, as in double auction markets (see Section~\ref{sec:mm}), the first closure operation in Definition~\ref{def:def2} is superfluous. Indeed, by \cite[Corollary~23.5.1]{roc70a}, $\rge\partial S_t=\dom\partial S_t^*$ where $S_t^*$ is the conjugate of $S_t$. Since, by \cite[Theorem~19.2]{roc70a}, $S_t^*$ is polyhedral, \cite[Theorem~23.10]{roc70a} implies $\dom\partial S_t^*=\dom S_t^*$ which is a polyhedral set and thus closed. Similarly, the set $\rge N_{D_t}$ is closed when $D$ is polyhedral.

\begin{theorem}\label{thm:def2}
If there is a strictly positive marginal price deflator, then
\[
C(S^\infty,D^\infty)\cap\M_+=\{0\}.
\]
On the other hand, if $S'$ is finite-valued and 
\[
\cl C(S^\infty,D^\infty)\cap\M_+=\{0\},
\]
then there is a strictly positive marginal price deflator. Moreover, in this case, the marginal price deflator can be chosen bounded.
\end{theorem}

\begin{proof}
See the appendix.
\qed\end{proof}

The following illustrates the above theorems with the exponential model from {\c{C}}etin and Rogers~\cite{cr7} where one has a cash account and one risky asset subject to illiquidity. 

\begin{example}
Let $J=\{0,1\}$, $D_t(\omega)=\reals^J$ and 
\[
S_t(x,\omega) = x^0 + \bar s_ t(\omega)\frac{e^{\alpha x^1}-1}{\alpha},
\]
where $\bar s=(\bar s_t)_{t=0}^T$ is a strictly positive stochastic market price process for the risky asset and $\alpha$ is a strictly positive scalar determining the degree of illiquidity for the risky asset. In this case, $S_t$ is differentiable and
\[
\nabla S_t(x,\omega)=(1,\bar s_t(\omega)e^{\alpha x^1}),
\]
so it follows that $\nabla S_t(0,\omega)=(1,\bar s_t(\omega))$ and 
\[
S'(x,\omega) = x^0+\bar s(\omega)\cdot x^1.
\]
The cost process $S'$ has the classical linear form so it follows from the closedness results of \cite{sch92} (see also \cite{ks1}) and Example~\ref{ex:numeraire} that $C(S',D')$ is closed whenever $C(S',D')\cap\M_+=\{0\}$. In this case, all the conditions in Theorem~\ref{thm:def1} become equivalent. Market price deflators are the positive adapted processes $y$ such that
\begin{align*}
E[y_{t+1}\,|\,\F_t]-y_t &= 0,\\
E[y_{t+1}\bar s_{t+1}\,|\,\F_t]-y_t\bar s_t &= 0.
\end{align*}
Thus, in the presence of cash account, the deflators become martingales. If $y$ is strictly positive, $y_T/y_0$ defines a strictly positive martingale density for the market price process $\bar s$. 

On the other hand, as long as the market price $\bar s$ is strictly positive, we get
\[
S^\infty_t(x,\omega)=
\begin{cases}
x^0 & \text{if $x^1\le 0$},\\
+\infty & \text{otherwise},
\end{cases}
\]
which is independent of $\bar s$. It is easily checked directly that $C(S^\infty,D^\infty)=\{c\in\M\,|\,\sum_{t=0}^Tc_t\le 0\}$ (see Example~\ref{ex:numeraire}) which is a closed set. Since $\rge\nabla S_t(\cdot,\omega)=\{1\}\times(0,+\infty)$, marginal price deflators are the positive adapted processes $y$ such that
\begin{align*}
E[y_{t+1}\,|\,\F_t]-y_t &= 0,\\
E[y_{t+1}s_{t+1}\,|\,\F_t]-y_ts_t &= 0
\end{align*}
for some strictly positive process $s$. In other words, any nonnegative martingale $y$ will be a marginal price deflator for $S$ and $D$.
\end{example}

\section{Conclusions}

Illiquidity effects give rise to market models where the set of freely available claim processes is nonconical. This paper suggested two generalizations of the no arbitrage condition for such models, a marginal and a scalable one. The two conditions were related to two notions of deflators that generalize martingale measures beyond classical perfectly liquid markets. 

Another practically important question is superhedging and pricing of contingent claims in nonconical market models. In classical perfectly liquid models as well as in conical models such as those in \cite{kab99,jn1,sch4}, superhedging is closely related to the deflators characterizing the no arbitrage or the no free lunch conditions. When moving to nonconical models, such as the one studied here, the situation changes. This is the topic of the follow-up paper \cite{pen8b}; see also \cite{pp8}, where the approach is further extended to a nonconical version of the currency market model of \cite{kab99}.

\section{Appendix}\label{sec:proofs}\label{sec:pf}

\begin{proof}[Proposition~\ref{prop:cones}]
If $\alpha>0$, we have $S'\le\alpha\star S$ and $D'\supset\alpha D$ so that
\begin{align*}
C(S',D') &= \{c\,|\,\exists x:\ x_t\in D'_t,\ S'_t(\Delta x_t)+c_t\le 0\}\\
&\supset\{c\,|\,\exists x:\ x_t\in\alpha D_t,\ \alpha\star S_t(\Delta x_t)+c_t\le 0\}\\
&= C(\alpha\star S,\alpha D)\\
&= \alpha C(S,D)
\end{align*}
so $\bigcup_{\alpha>0}\alpha C(S,D)\subset C(S',D')$. On the other hand, if $c\in C(S',D')$ there is an $x$ such that $S'_t(\Delta x_t)+c_t \le 0$. Let $x^\alpha_t(\omega)$ be the Euclidean projection of $x_t(\omega)$ on $\alpha D_t(\omega)$. By \cite[Theorem~14.37]{rw98}, this defines an adapted process $x^\alpha$ that, by \cite[Proposition~4.9]{rw98}, converges to $x$ almost surely. Defining
\[
c^\alpha_t = S'_t(\Delta x_t)-\alpha\star S_t(\Delta x^\alpha_t)+c_t
\]
we have
\[
\alpha\star S_t(\Delta x^\alpha_t)+c^\alpha_t \le 0
\]
so that $c^\alpha(x)\in\alpha C(S,D)$.
Since $S_t(\cdot,\omega)$ are finite by assumption, \cite[Theorem~23.1]{roc70a} implies that $c^\alpha$ converges almost surely to $c$ as $\alpha\upto\infty$. Thus, $C(S',D')\subset\cl\bigcup_{\alpha>0}\alpha C(S,D)$.

To prove the second claim, we first note that for $\alpha>0$, we have $D^\infty\subset\alpha D$, $S^\infty\ge\alpha\star S$ so that
\begin{align*}
C(S^\infty,D^\infty) &= \{c\,|\,\exists x:\ x_t\in D_t^\infty,\ S^\infty_t(\Delta x_t)+c_t\le 0\}\\
&\subset\{c\,|\,\exists x:\ x_t\in\alpha D_t,\ \alpha\star S_t(\Delta x_t)+c_t\le 0\}\\
&= C(\alpha\star S,\alpha D)\\
&= \alpha C(S,D),
\end{align*}
so $C(S^\infty,D^\infty)\subset\bigcap_{\alpha>0}\alpha C(S,D)$. On the other hand, under the extra assumptions on $S$ and $D$, we get
\begin{align*}
\alpha C(S,D) &=\{c\,|\, \exists x:\ x_t\in\alpha D_t,\ \alpha S_t\left(\frac{\Delta x_t}{\alpha}\right) + c_t\le 0\}\\
&\subset\{c\,|\, \exists x:\ x_t\in D^\infty_t+\alpha a\uball,\ S^\infty_t\left(\Delta x_t\right) -\alpha a + c_t\le 0\}\\
&=\{c + \alpha a \,|\, \exists x:\ x_t\in D^\infty_t+\alpha a\uball,\ S^\infty_t\left(\Delta x_t\right) + c_t\le 0\}\\
&\subset\{c + \alpha a \,|\, \exists x:\ x_t\in D^\infty_t,\ \inf_{z\in 2\alpha a\uball}S^\infty_t\left(\Delta x_t+z\right) + c_t\le 0\}.
\end{align*}
Since $S\ge S^\infty-a$, by assumption, the finiteness of $S$ implies that of $S^\infty$. Since $S^\infty$ is sublinear, $S^\infty_t(\cdot,\omega)$ will then be Lipschitz continuous and the Lipschitz constant $L(\omega)$ can be chosen measurable. We get
\begin{align*}
\alpha C(S,D) &\subset\{c + \alpha a \,|\, \exists x:\ x_t\in D^\infty_t,\ S^\infty_t\left(\Delta x_t\right) -2\alpha aL + c_t\le 0\}\\
&=C(S^\infty,D^\infty)+\alpha(a+2aL).
\end{align*}
So for every $c\in\bigcap_{\alpha>0}\alpha C(S,D)$ and $\alpha>0$ there is a $c^\alpha\in C(S^\infty,D^\infty)$ such that $c=c^\alpha+\alpha(a+2aL)$. As $\alpha\downto 0$, $c^\alpha$ converges almost surely to $c$. Thus $c\in\cl C(S^\infty,D^\infty)$.
\qed\end{proof}

To prove Theorems~\ref{thm:def1} and \ref{thm:def2}, we will use functional analytic techniques much as e.g.\ in \cite{sch92,fs4,ds6}. Due to possible nonlinearities, however, our model requires a bit more convex analysis than traditional linear models. In particular, a major role is played by the theory of {\em normal integrands} (see e.g.~\cite{roc68,roc76,rw98}), which was the reason for including the measurability and closedness conditions in Definitions~\ref{pp} and \ref{pcp}.

Given a cost process $S$ and a constraint process $D$, consider the function $\sigma_{C(S,D)}:\M\to\ereals$ defined by\footnote{Here and in what follows, we define the expectation $E\varphi$ of an arbitrary measurable function $\varphi$ by setting $E\varphi=-\infty$ unless the negative part of $\varphi$ is integrable. The expectation is then a well-defined extended real number for any measurable function.}
\[
\sigma_{C(S,D)}(y) = \sup\left\{\left.E\sum_{t=0}^Tc_t y_t\,\left.\right.\right|\,c\in C(S,D)\right\}.
\]
The function $\sigma_{C(S,D)}$ is nonnegative since $0\in C(S,D)$. Also, since $C(S,D)$ contains all nonpositive claim processes, the effective domain 
\[
\dom\sigma_{C(S,D)} = \{y\in\M\,|\,\sigma_{C(S,D)}(y)<\infty\}
\]
of $\sigma_{C(S,D)}$ is contained in $\M_+$. In the terminology of microeconomic theory, $\sigma_{C(S,D)}$ is called the {\em profit function} associated with the ``production set'' $C(S,D)$; see e.g.\ Aubin~\cite{aub79} or Mas-Collel, Whinston and Green~\cite{mwg95}. 

We will derive an expression for $\sigma_{C(S,D)}$ in terms of $S$ and $D$. This will involve the space $\N^1$ of $\reals^J$-valued adapted integrable processes $v=(v_t)_{t=0}^T$ and the integral functionals 
\[
v_t\mapsto E(y_tS_t)^*(v_t)\quad\text{and}\quad v_t\mapsto E\sigma_{D_t}(v_t)
\]
associated with the normal integrands
\begin{align*}
(y_tS_t)^*(v,\omega) &:= \sup_{x\in\reals^J}\{x\cdot v - y_t(\omega)S_t(x,\omega)\}
\intertext{and}
\sigma_{D_t(\omega)}(v) &:= \sup_{x\in\reals^J}\{x\cdot v\,|\, x\in D_t(\omega)\}.
\end{align*}
That the above expressions do define normal integrands follows from \cite[Theorem~14.50]{rw98}. Since $S_t(0,\omega)=0$ and $0\in D_t(\omega)$ for every $t$ and $\omega$, the functions $(y_tS_t)^*$ and $\sigma_{D_t}$ are nonnegative.

Let $\M^1$ and $\M^\infty$ be the spaces of integrable and essentially bounded, respectively, real-valued adapted processes. The bilinear form
\[
(c,y)\mapsto E\sum_{t=0}^Tc_ty_t
\]
puts the spaces $\M^1$ and $\M^\infty$ in separating duality; see \cite{roc74}. 

We will say that a cost process $S$ is {\em integrable} if the functions $S_t(x,\cdot)$ are integrable for every $t=0,\ldots,T$ and $x\in\reals^J$. In the classical linear case $S_t(x,\omega)=s_t(\omega)\cdot x$, integrability means that price vectors $s_t$ are integrable. 

\begin{lemma}\label{lem:pid}
For $y\in\M_+$, 
\[
\sigma_{C(S,D)}(y) \le \inf_{v\in\N^1}\left\{\sum_{t=0}^T E(y_t S_t)^*(v_t) + \sum_{t=0}^{T-1}E\sigma_{D_t}(E[\Delta v_{t+1}|\F_t])\right\},
\]
while $\sigma_{C(S,D)}(y)=+\infty$ for $y\notin\M_+$. If $S$ is integrable then equality holds and the infimum is attained for every $y\in\M^\infty_+$.
\end{lemma}

\begin{proof}
Only the case $y\in\M_+$ requires proof so assume that. Let $v\in\N^1$ be arbitrary. We have
\begin{align}
\sigma_{C(S,D)}(y) &= \sup_{x\in\N_0,c\in\M}\left\{\left.E\sum_{t=0}^Ty_tc_t\,\right|\, S_t(\Delta x_t)+c_t\le 0,\ x_t\in D_t\right\} \nonumber\\
&= \sup_{x\in\N_0}\left\{\left.E\left[-\sum_{t=0}^Ty_tS_t(\Delta x_t)\right]\,\right|\, x_t\in D_t\right\} \nonumber\\
&= \sup_{x\in\N_0}\left\{\left.E\left[\sum_{t=0}^T(\Delta x_t\cdot v_t-y_tS_t(\Delta x_t))-\sum_{t=0}^T\Delta x_t\cdot v_t\right]\,\right|\, x_t\in D_t\right\}\nonumber \\
&\le\sup_{x\in\N_0}\left\{\left.E\left[\sum_{t=0}^T(y_tS_t)^*(v_t) + \sum_{t=0}^{T-1}x_t\cdot\Delta v_{t+1}\right]\,\right|\, x_t\in D_t\right\}\nonumber\\
&= E\sum_{t=0}^T(y_tS_t)^*(v_t) + \sup_{x\in\N_0}\left\{\left.E\sum_{t=0}^{T-1}x_t\cdot\Delta v_{t+1}\,\right|\, x_t\in D_t\right\}.\label{ie0}
\end{align}

Assume first that the last term in \eqref{ie0} if finite. Let $\eps>0$ be arbitrary and let $x'\in\N_0$ be such that $x'_t\in D_t$ and 
\begin{equation}\label{ie1}
E\sum_{t=0}^{T-1}x'_t\cdot\Delta v_{t+1} \ge \sup_{x\in\N_0}\left\{\left.E\sum_{t=0}^{T-1}x_t\cdot\Delta v_{t+1}\,\right|\, x_t\in D_t\right\} - \eps.
\end{equation}
Since $0\in D_t$, the supremum is nonnegative and thus, the negative part of $\sum_{t=0}^{T-1}x'_t\cdot\Delta v_{t+1}$ must be integrable. Let $A_t^\nu=\{\omega\,|\, |x'_t(\omega)|\le\nu\}$ and $x^\nu_t=x'_t\chi_{A_t^\nu}$. We then have $x^\nu\to x'$ almost surely as $\nu\to\infty$, so by Fatou's lemma, there is a $\bar\nu$ such that
\begin{equation}\label{ie2}
E\sum_{t=0}^{T-1}x^{\bar\nu}_t\cdot\Delta v_{t+1}\ge E\sum_{t=0}^{T-1}x'_t\cdot\Delta v_{t+1} - \eps.
\end{equation}
Since $x^{\bar\nu}$ is bounded and since $x^{\bar\nu}_t\in D_t$, we get
\begin{align}
E\sum_{t=0}^{T-1}x^{\bar\nu}_t\cdot\Delta v_{t+1} &= E\sum_{t=0}^{T-1}x^{\bar\nu}_t\cdot E[\Delta v_{t+1}\,|\,\F_t]\nonumber\\
&\le E\sum_{t=0}^{T-1}\sigma_{D_t}\left(E[\Delta v_{t+1}\,|\,\F_t]\right).\label{ie3}
\end{align}
Combining \eqref{ie0}--\eqref{ie3} we get
\[
\sigma_{C(S,D)}(y) \le \sum_{t=0}^TE(y_tS_t)^*(v_t) + \sum_{t=0}^{T-1}E\sigma_{D_t}\left(E[\Delta v_{t+1}\,|\,\F_t]\right) + 2\varepsilon.
\]
Since $v\in\N^1$ and $\eps>0$ were arbitrary, we get that the first claim holds when the last term in \eqref{ie0} is finite.  When the last term in \eqref{ie0} equals $+\infty$, there is an $x'\in\N_0$ such that $x'_t\in D_t$ and 
\[
E\sum_{t=0}^{T-1}x'_t\cdot\Delta v_{t+1} \ge 1/\eps.
\]
A similar argument as above then shows that 
\[
\sum_{t=0}^{T-1}E\sigma_{D_t}\left(E[\Delta v_{t+1}\,|\,\F_t]\right) = +\infty
\]
which completes the proof of the first claim.

To prove the second claim, it suffices to show that when $S$ is integrable and $y\in\M^\infty_+$, the right side in the inequality equals the support function $\sigma_{C^1_\infty(S,D)}:\M^\infty\to\ereals$ of the set
\[
C^1_ \infty(S,D) = \{c\in\M^1\,|\,\exists x\in\N^\infty_0:\ x_t\in D_t,\ S_t(\Delta x_t)+c_t\le 0,\ t=0,\ldots,T\},
\]
where $\N^\infty_0\subset\N_0$ is the space of essentially bounded portfolio processes with $x_T=0$. Indeed, since $C^1_\infty(S,D)\subset C(S,D)$ we have $\sigma_{C^1_\infty(S,D)}\le\sigma_{C(S,D)}$. 

When $S$ is integrable, we have that $\omega\mapsto S_t(z(\omega),\omega)$ is integrable for every $z\in L^\infty(\Omega,\F,P;\reals^J)$. Indeed (see \cite[Theorem~3K]{roc76}), if $\|z\|_{L^\infty}\le r$, there is a finite set of points $x^i\in\reals^J$ $i=1,\ldots,n$ whose convex combination contains the ball $r\uball$. By convexity, $S_t(z(\omega),\omega)\le\sup_{i=1,\ldots,n}S_t(x^i,\omega)$, where the right hand side is integrable by assumption. It follows that for $y\in\M^\infty_+$
\begin{align*}
\sigma_{C^1_\infty(S,D)}(y) &= \sup_{x\in\N^\infty_0,\ c\in\M^1}\{E\sum_{t=0}^Tc_ty_t\,|\, x_t\in D_t,\ S_t(\Delta x_t)+c_t\le 0\}\\
&= \sup_{x\in\N^\infty_0}\{E\sum_{t=0}^T-(y_tS_t)(\Delta x_t)\,|\, x_t\in D_t\}\\
&= \sup_{x\in\N^\infty_0}E\left\{-\sum_{t=0}^T(y_tS_t)(\Delta x_t) - \sum_{t=0}^{T-1}\delta_{D_t}(x_t)\right\},
\end{align*}
where $\delta_{D_t}$ denotes the normal integrand defined by $\delta_{D_t(\omega)}(x)=0$ if $x\in D_t(\omega)$ and $\delta_{D_t(\omega)}(x)=+\infty$ otherwise. When $S$ is integrable, we can write
\begin{equation}\label{eq:fr}
\sigma_{C^1_\infty(S,D)}(y) = -\inf_{x\in\N^\infty_0}\{h(x)+k(Ax)\},
\end{equation}
where $h:\N^\infty_0\to\ereals$, $k:\N^\infty\to\reals$ and $A:\N^\infty_0\to\N^\infty$ are defined by 
\begin{align*}
h(x) &= \sum_{t=0}^{T-1}E\delta_{D_t}(x_t),\\
k(x) &= \sum_{t=0}^TE(y_tS_t)(x_t),\\
Ax &= (x_0,x_1-x_0,\ldots,-x_{T-1}).
\end{align*}
The bilinear form $(x,v)\mapsto E\sum_{t=0}^{T-1}x_t\cdot v_t$ puts $\N^\infty_0$ and $\N^1_0$ in separating duality. We pair $\N^\infty$ and $\N^1$ similarly. The expression \eqref{eq:fr} then fits the Fenchel-Rockafellar duality framework; see \cite{roc67} or Examples~11 and 11$'$ of \cite{roc74}. When $y\in\M^\infty_+$, the integrability of $S$ implies that $k$ is finite on all of $\N^\infty$ and then, by \cite[Theorem~22]{roc74}, it is continuous with respect to the Mackey topology. Theorems~1 and 3 of \cite{roc67} (or Theorem~17 of \cite{roc74}) then give,
\begin{equation}\label{dual}
\sigma_{C^1_\infty(S,D)}(y) = \min_{v\in\N^1}\{k^*(v) + h^*(-A^*v)\},
\end{equation}
where the minimum is attained, $A^*:\N^1\to\N^1_0$ is the adjoint of $A$ and $k^*:\N^1\to\ereals$ and $h^*:\N^1_0\to\ereals$ are the conjugates of $k$ and $h$, respectively. It is not hard to check that
\[
A^*v = -\left(E[\Delta v_1|\F_0],\ldots,E[\Delta v_T|\F_{T-1}]\right).
\]
Writing
\[
k(x)=k_0(x_0)+\ldots+k_T(x_T),
\]
where $k_t:L^\infty(\Omega,\F_t,P;\reals^J)\to\reals$ is given by
\[
k_t(x_t)=E[y_t(\omega)S_t(x_t(\omega),\omega)],
\]
we get
\[
k^*(v) = k_0^*(v_0)+\ldots + k_T^*(v_T),
\]
where, by \cite[Theorem~21]{roc74}, $k_t^*(v_t) = E(y_tS_t)^*(v_t(\omega),\omega)$. By \cite[Theorem~21]{roc74} again, 
\[
h^*(v) = E\sum_{t=0}^{T-1}\sigma_{D_t}(v_t).
\]
Plugging the above expressions for $A^*$, $k^*$ and $h^*$ in \eqref{dual} gives the desired expression. 
\qed\end{proof}


\begin{proof}[Theorem~\ref{thm:def1}]
We begin by showing that a strictly positive $y\in\M_+$ is a market price deflator iff 
\begin{equation}\label{ie}
\exists\ v\in\N^1:\quad\sum_{t=0}^T E(y_tS'_t)^*(v_t) + \sum_{t=0}^{T-1}E\sigma_{D'_t}\left(E[\Delta v_{t+1}|\F_t]\right)\le 0.
\end{equation}
Since $S'_t(0,\omega)=0$ and $0\in D'_t(\omega)$ for every $\omega\in\Omega$, we have $(y_tS'_t)^*(v,\omega)\ge 0$ and $\sigma_{D'_t(\omega)}\left(v\right)\ge 0$ for every $v\in\reals^J$ and $\omega\in\Omega$ so the inequality in \eqref{ie} means that
\[
v_t(\omega)\in\argmin (y_tS'_t)^*(\cdot,\omega)\quad\text{and}\quad E[\Delta v_{t+1}|\F_t](\omega)\in\argmin\sigma_{D'_t(\omega)}.
\]
By \cite[Theorem~23.5]{roc70a},
\[
\argmin (y_tS'_t)^*(\cdot,\omega) = y_t(\omega)\partial S'_t(0,\omega),
\]
and
\[
\argmin\sigma_{D'_t(\omega)} = \partial\delta_{D'_t(\omega)}(0),
\]
where $\partial S'_t(0,\omega)=\partial S_t(0,\omega)$ and $\partial\delta_{D'_t(\omega)}(0)=N_{D_t(\omega)}(0)$. Defining $s=v/y$, we see that $y$ is a market price deflator in the sense of Definition~\ref{def:def1} iff \eqref{ie} holds.

Applying the first part of Lemma~\ref{lem:pid} to $S'$ and $D'$, we thus get that a strictly positive marginal price deflator $y$ satisfies $\sigma_{C(S',D')}(y)\le 0$, or equivalently,
\[
E\sum_{t=0}^Tc_ty_t \le 0\quad \forall c\in C(S',D').
\]
On the other hand, strict positivity of $y$ means that 
\[
E\sum_{t=0}^Tc_ty_t \le 0\quad \forall c\in\M_+,
\]
so we must have $C(S',D')\cap\M_+=\{0\}$ whenever a strictly positive market price deflator exists.

In proving the second claim, we may assume that $S'$ is integrable. Indeed, when $S'$ is finite-valued we can define an equivalent measure $\tilde P$ with the bounded strictly positive density $d\tilde P/dP =\psi := e^{-\varphi}/Ee^{-\varphi}$, where
\[
\varphi(\omega) = \max_{t=0,\ldots,T}\sup_{x\in\uball} S_t'(x,\omega)
\]
is finite and nonnegative. By equivalence, the condition $\cl C(S',D')\cap\M_+=\{0\}$ holds under $P$ iff it holds under $\tilde P$. Also, $S'$ is integrable under $\tilde P$ since for every nonzero $x\in\reals^J$
\[
E^{\tilde P} |S'_t(x,\omega)| = |x|E^{\tilde P} |S'_t(x/|x|,\omega)| \le |x|E^{\tilde P}\varphi = \frac{|x|}{Ee^{-\varphi}}E\varphi e^{-\varphi} <\infty.
\]
Moreover, if $\tilde y$ is a bounded strictly positive market price deflator under $\tilde P$, then the bounded strictly positive process $y_t := E[\psi\,|\,\F_t]\tilde y_t$ is a market price deflator under $P$. Indeed, if there is a market price process $s\in\partial S(0)$ such that 
\[
E^{\tilde P}[y_{t+1}\tilde s_{t+1}\,|\F_t]-y_t\tilde s_t\in N_{D_t}(0),
\]
then
\begin{align*}
E[y_{t+1}s_{t+1}\,|\,\F_t]-y_ts_t &= E\left[E\left[\psi\,|\,\F_{t+1}\right]\tilde y_{t+1}s_{t+1}\,|\,\F_t\right]-E\left[\psi\,|\,\F_t\right]\tilde y_ts_t\\
&= E\left[\psi\,|\,\F_t\right]E^{\tilde P}\left[\tilde y_{t+1}s_{t+1}\,|\,\F_t\right]-E\left[\psi\,|\,\F_t\right]\tilde y_ts_t\\
&\in E\left[\psi\,|\,\F_t\right]N_{D_t}(0)\\
&=N_{D_t}(0),
\end{align*}
where the second equality holds by \cite[Proposition A.12]{fs4} and the last since $N_{D_t}(0)$ is an $\F_t$-measurable cone. We would thus have that $y=(y_t)_{t=0}^T$ is a bounded strictly positive market price deflator under $P$, so without loss of generality, we may assume that $S'$ is integrable. 

Note that 
\[
\cl C(S',D')\cap\M_+=\{0\} \implies\cl C^1(S',D')\cap\M_+^1=\{0\},
\]
where $C^1(S',D')=C(S',D')\cap\M^1$. By the Kreps-Yan theorem (see e.g.\ \cite{jns5} or \cite{rok5}), the latter condition implies the existence of a $y\in\M^\infty$ such that 
\begin{align}
E\sum_{t=0}^Tc_ty_t &> 0\quad \forall c\in\M^1_+\setminus\{0\},\label{ky1}\\
E\sum_{t=0}^Tc_ty_t &\le 0\quad \forall c\in C^1(S',D').\label{ky2}
\end{align}
The first condition means that $y$ is almost surely strictly positive while the second means that $\sigma_{C^1(S',D')}(y)\le 0$. We clearly have $\sigma_{C^1_\infty(S',D')}\le\sigma_{C^1(S',D')}\le\sigma_{C(S',D')}$, where $C^1_\infty(S',D')$ is as in the proof of Lemma~\ref{lem:pid}, where the equality $\sigma_{C^1_\infty(S',D')}=\sigma_{C(S',D')}$ was established under the integrability condition. Thus, under integrability, \eqref{ky2} holds iff $\sigma_{C(S',D')}(y)\le 0$ which by Lemma~\ref{lem:pid}, implies \eqref{ie} and thus that $y$ is a strictly positive market price deflator.
\qed\end{proof}

\begin{proof}[Theorem~\ref{thm:def2}]
This is analogous to the proof of Theorem~\ref{thm:def1}. It suffices to replace  $S'$ and $D'$ by $S^\infty$ and $D^\infty$, respectively, and to show that when $y$ is strictly positive,
\[
\argmin (y_tS^\infty_t)^*(\cdot,\omega) = y_t(\omega)\cl\rge\partial S_t(\cdot,\omega)
\]
and
\[
\argmin\sigma_{D^\infty_t(\omega)} = \cl\rge N_{D_t(\omega)}.
\]
Applying \cite[Theorem~23.5]{roc70a} to the functions $y_t(\omega)S^\infty_t(\cdot,\omega)$ and $\delta_{D_t^\infty(\omega)}$, we get
\[
\argmin (y_tS^\infty_t)^*(\cdot,\omega) = y_t(\omega)\partial S^\infty_t(0,\omega)
\]
and
\[
\argmin\sigma_{D^\infty_t(\omega)} = \partial\delta_{D^\infty_t(\omega)}(0).
\]
By Theorem~13.3 of \cite{roc70a}, $S^\infty(\cdot,\omega)$ is the support function of $\dom S_t^*(\cdot,\omega)$, which implies that $\partial S_t^\infty(0,\omega)=\cl\dom S_t^*(\cdot,\omega)$. By Theorem~23.4 and Corollary~23.5.1 of \cite{roc70a}, 
\[
\cl\dom S_t^*(\cdot,\omega)=\cl\dom\partial S_t^*(\cdot,\omega)=\cl\rge\partial S_t(\cdot,\omega). 
\]
This gives the desired expression for $\argmin (y_tS^\infty_t)^*(\cdot,\omega)$. The expression for $\argmin\sigma_{D^\infty_t(\omega)}$ follows similarly by noting that $\delta_{D_t^\infty}$ is the horizon function of $\delta_{D_t(\omega)}$ and that $N_{D_t(\omega)}(x)=\partial\delta_{D_t(\omega)}(x)$ for every $x\in\reals^J$.
\qed\end{proof}


\bibliographystyle{plain}
\bibliography{sp.bib}   

\begin{thebibliography}{10}

\bibitem{ass}
A.~Alfonsi, A.~Schied, and A.~Schulz.
\newblock Constrained portfolio liquidation in a limit order book model.
\newblock {\em Preprint}, 2007.

\bibitem{ac}
R.~Almgren and N.~Chriss.
\newblock Optimal execution of portfolio transactions.
\newblock {\em Journal of Risk}, 3:5--39, 2000.

\bibitem{at7}
F.~Astic and N.~Touzi.
\newblock No arbitrage conditions and liquidity.
\newblock {\em J. Math. Econom.}, 43:692--708, 2007.

\bibitem{aub79}
J.-P. Aubin.
\newblock {\em Mathematical methods of game and economic theory}, volume~7 of
  {\em Studies in Mathematics and its Applications}.
\newblock North-Holland Publishing Co., Amsterdam, 1979.

\bibitem{bb4}
P.~Bank and D.~Baum.
\newblock Hedging and portfolio optimization in financial markets with a large
  trader.
\newblock {\em Math. Finance}, 14(1):1--18, 2004.

\bibitem{bdes98}
H.~B\"uhlmann, F.~Delbaen, P.~Embrechts, and A.~Shiryaev.
\newblock On {Esscher} transforms in discrete finance models.
\newblock {\em ASTIN Bulletin}, 28:171--186, 1998.

\bibitem{cjp4}
U.~{\c{C}}etin, R.~A. Jarrow, and P.~Protter.
\newblock Liquidity risk and arbitrage pricing theory.
\newblock {\em Finance Stoch.}, 8(3):311--341, 2004.

\bibitem{cr7}
U.~{\c{C}}etin and L.~C.~G. Rogers.
\newblock Modelling liquidity effects in discrete time.
\newblock {\em Mathematical Finance}, 17(1):15--29, 2007.

\bibitem{ck92}
J.~Cvitani{\'c} and I.~Karatzas.
\newblock Convex duality in constrained portfolio optimization.
\newblock {\em Ann. Appl. Probab.}, 2(4):767--818, 1992.

\bibitem{ds6}
F.~Delbaen and W.~Schachermayer.
\newblock {\em The Mathematics of Arbitrage}.
\newblock Springer Finance. Springer-Verlag, Berlin Heidelberg, 2006.

\bibitem{dr91}
J.~C. Dermody and R.~T. Rockafellar.
\newblock Cash stream valuation in the face of transaction costs and taxes.
\newblock {\em Math. Finance}, 1(1):31--54, 1991.

\bibitem{dr95}
J.~C. Dermody and R.~T. Rockafellar.
\newblock Tax basis and nonlinearity in cash stream valuation.
\newblock {\em Math. Finance}, 5(2):97--119, 1995.

\bibitem{est4}
I.~V. Evstigneev, K.~Sch{\"u}rger, and M.~I. Taksar.
\newblock On the fundamental theorem of asset pricing: random constraints and
  bang-bang no-arbitrage criteria.
\newblock {\em Math. Finance}, 14(2):201--221, 2004.

\bibitem{fs4}
H.~F{\"o}llmer and A.~Schied.
\newblock {\em Stochastic finance}, volume~27 of {\em de Gruyter Studies in
  Mathematics}.
\newblock Walter de Gruyter \& Co., Berlin, extended edition, 2004.
\newblock An introduction in discrete time.

\bibitem{jk95a}
E.~Jouini and H.~Kallal.
\newblock Martingales and arbitrage in securities markets with transaction
  costs.
\newblock {\em J. Econom. Theory}, 66(1):178--197, 1995.

\bibitem{jn1}
E.~Jouini and C.~Napp.
\newblock Arbitrage and investment opportunities.
\newblock {\em Finance Stoch.}, 5(3):305--325, 2001.

\bibitem{jns5}
E.~Jouini, C.~Napp, and W.~Schachermayer.
\newblock Arbitrage and state price deflators in a general intertemporal
  framework.
\newblock {\em J. Math. Econom.}, 41(6):722--734, 2005.

\bibitem{kab99}
Yu.~M. Kabanov.
\newblock Hedging and liquidation under transaction costs in currency markets.
\newblock {\em Finance and Stochastics}, 3(2):237--248, 1999.

\bibitem{krs3}
Yu.~M. Kabanov, M.~R{\'a}sonyi, and Ch. Stricker.
\newblock On the closedness of sums of convex cones in {$L\sp 0$} and the
  robust no-arbitrage property.
\newblock {\em Finance Stoch.}, 7(3):403--411, 2003.

\bibitem{ks1}
Yu.~M. Kabanov and Ch. Stricker.
\newblock A teachers' note on no-arbitrage criteria.
\newblock In {\em S\'eminaire de Probabilit\'es, XXXV}, volume 1755 of {\em
  Lecture Notes in Math.}, pages 149--152. Springer, Berlin, 2001.

\bibitem{km6}
K.~Kaval and I.~Molchanov.
\newblock Link-save trading.
\newblock {\em J. Math. Economics}, 42:710--728, 2006.

\bibitem{ks72}
A.~Kraus and H.~R. Stoll.
\newblock Price impacts of block trading on the {New York} stock exchange.
\newblock {\em Journal of Finance}, 27(3):569--588, 1972.

\bibitem{ku7}
P.~Krokhmal and S.~Uryasev.
\newblock A sample-path approach to optimal position liquidation.
\newblock {\em Ann. Oper. Res.}, 152(1):193--225, 2007.

\bibitem{kuh6}
C.~K\"uhn.
\newblock Optimal investment in financial markets with different liquidity
  effects.
\newblock {\em Submitted for publication}, 2006.

\bibitem{mp8}
P.~Malo and T.~Pennanen.
\newblock Marginal prices of market orders: convexity and statistic.
\newblock {\em Manuscript}.

\bibitem{mwg95}
A.~Mas-Collel, M.D.\ Whinston, and J.R.\ Green.
\newblock {\em Microeconomic Theory}.
\newblock Oxford University Press, New York, 1995.

\bibitem{nap3}
C.~Napp.
\newblock The {D}alang-{M}orton-{W}illinger theorem under cone constraints.
\newblock {\em J. Math. Econom.}, 39(1-2):111--126, 2003.
\newblock Special issue on equilibrium with asymmetric information.

\bibitem{pen6}
T.~Pennanen.
\newblock Nonlinear price processes.
\newblock {\em Preprint, http://math.tkk.fi/$\sim$teemu}, 2006.

\bibitem{pen7}
T.~Pennanen.
\newblock Free lunches and martingales in convex markets.
\newblock {\em Preprint available at http://math.tkk.fi/$\sim$teemu}, 2007.

\bibitem{pen8b}
T.~Pennanen.
\newblock Superhedging in illiquid markets.
\newblock {\em Submitted}, 2008.

\bibitem{pp8}
T.~Pennanen and I.~Penner.
\newblock Hedging of claims with physical delivery under convex transaction
  costs.
\newblock {\em Submitted}, 2008.

\bibitem{pt99}
H.~Pham and N.~Touzi.
\newblock The fundamental theorem of asset pricing with cone constraints.
\newblock {\em J. Math. Econom.}, 31(2):265--279, 1999.

\bibitem{ps98}
E.~Platen and M.~Schweizer.
\newblock On feedback effects from hedging derivatives.
\newblock {\em Math. Finance}, 8(1):67--84, 1998.

\bibitem{roc67}
R.~T. Rockafellar.
\newblock Duality and stability in extremum problems involving convex
  functions.
\newblock {\em Pacific J. Math.}, 21:167--187, 1967.

\bibitem{roc68}
R.~T. Rockafellar.
\newblock Integrals which are convex functionals.
\newblock {\em Pacific J. Math.}, 24:525--539, 1968.

\bibitem{roc70a}
R.~T. Rockafellar.
\newblock {\em Convex analysis}.
\newblock Princeton Mathematical Series, No. 28. Princeton University Press,
  Princeton, N.J., 1970.

\bibitem{roc74}
R.~T. Rockafellar.
\newblock {\em Conjugate duality and optimization}.
\newblock Society for Industrial and Applied Mathematics, Philadelphia, Pa.,
  1974.
\newblock Lectures given at the Johns Hopkins University, Baltimore, Md., June,
  1973, Conference Board of the Mathematical Sciences Regional Conference
  Series in Applied Mathematics, No. 16.

\bibitem{roc76}
R.~T. Rockafellar.
\newblock Integral functionals, normal integrands and measurable selections.
\newblock In {\em Nonlinear operators and the calculus of variations (Summer
  School, Univ. Libre Bruxelles, Brussels, 1975)}, pages 157--207. Lecture
  Notes in Math., Vol. 543. Springer, Berlin, 1976.

\bibitem{rw98}
R.~T. Rockafellar and R.~J.-B. Wets.
\newblock {\em Variational analysis}, volume 317 of {\em Grundlehren der
  Mathematischen Wissenschaften [Fundamental Principles of Mathematical
  Sciences]}.
\newblock Springer-Verlag, Berlin, 1998.

\bibitem{rogsin6}
L.~C.~G. Rogers and S.~Singh.
\newblock Modelling liquidity and its effects on price.
\newblock {\em Preprint}, 2006.

\bibitem{rok2}
D.~B. Rokhlin.
\newblock A criterion for the nonexistence of the asymptotic free lunch in a
  finite-dimensional market under convex portfolio constraints and convex
  transaction costs.
\newblock {\em Sib. Zh. Ind. Mat.}, 5(1):133–--144, 2002.
\newblock (in Russian).

\bibitem{rok5b}
D.~B. Rokhlin.
\newblock An extended version of the {D}alang-{M}orton-{W}illinger theorem
  under convex portfolio constraints.
\newblock {\em Theory Probab. Appl.}, 49(3):429--443, 2005.

\bibitem{rok5}
D.~B. Rokhlin.
\newblock The {K}reps-{Y}an theorem for {$L\sp \infty$}.
\newblock {\em Int. J. Math. Math. Sci.}, (17):2749--2756, 2005.

\bibitem{rok7}
D.~B. Rokhlin.
\newblock Martingale selection problem and asset pricing in finite discrete
  time.
\newblock {\em Electron. Comm. Probab.}, 12:1--8, 2007.

\bibitem{sch92}
W.~Schachermayer.
\newblock A {H}ilbert space proof of the fundamental theorem of asset pricing
  in finite discrete time.
\newblock {\em Insurance Math. Econom.}, 11(4):249--257, 1992.

\bibitem{sch4}
W.~Schachermayer.
\newblock The fundamental theorem of asset pricing under proportional
  transaction costs in finite discrete time.
\newblock {\em Math. Finance}, 14(1):19--48, 2004.

\end{thebibliography}

\end{document}